\documentclass[twocolumn,showpacs,preprintnumbers,amsmath,amssymb,prl]{revtex4-1}
\usepackage{amsmath}
\usepackage{amsfonts}
\usepackage{amsthm}
\usepackage{graphics}
\usepackage{graphicx}
\usepackage{subfigure}
\usepackage{amssymb}
\usepackage{color}
\usepackage{url}
\usepackage[abs]{overpic}
\usepackage[usenames,dvipsnames]{xcolor}

\begin{document}

\title{Statistical theory of reversals in two-dimensional confined
turbulent flows}

\author{Vishwanath Shukla}
\email{research.vishwanath@gmail.com}
\affiliation{Laboratoire de Physique Statistique, 
\'{E}cole Normale Sup{\'e}rieure, 
PSL Research University;
UPMC Univ Paris 06, Sorbonne Universit\'{e}s;
Universit\'{e} Paris Diderot, Sorbonne Paris-Cit\'{e};
CNRS;
24 Rue Lhomond, 75005 Paris, France
}
\author{Stephan Fauve}
\email{fauve@lps.ens.fr}
\affiliation{Laboratoire de Physique Statistique, 
\'{E}cole Normale Sup{\'e}rieure, 
PSL Research University;
UPMC Univ Paris 06, Sorbonne Universit\'{e}s;
Universit\'{e} Paris Diderot, Sorbonne Paris-Cit\'{e};
CNRS;
24 Rue Lhomond, 75005 Paris, France
}\author{Marc Brachet} 
\email{brachet@physique.ens.fr}
\affiliation{Laboratoire de Physique Statistique, 
\'{E}cole Normale Sup{\'e}rieure, 
PSL Research University;
UPMC Univ Paris 06, Sorbonne Universit\'{e}s;
Universit\'{e} Paris Diderot, Sorbonne Paris-Cit\'{e};
CNRS;
24 Rue Lhomond, 75005 Paris, France
}
\date{\today}
\begin{abstract}
It is shown that the Truncated Euler Equations, {\emph i.e.} a finite set of ordinary differential equations for the amplitude 
of the large-scale modes, can correctly describe the complex transitional dynamics that occur within the turbulent regime of a confined $2$D Navier-Stokes flow with bottom friction and a spatially periodic forcing. In particular, the random reversals of the large scale circulation on the turbulent background involve bifurcations of the probability distribution function of the large-scale circulation velocity that are described by the related microcanonical distribution which displays transitions from gaussian to bimodal and broken ergodicity. A minimal $13$-mode model reproduces these results.
\end{abstract}
\pacs{47.27.-i, 47.27.E-,47.27.De}
\keywords{turbulence; bifurcations; absolute equilibrium}
\maketitle

The formation of large scale coherent structures is widely observed in atmospheric 
and oceanic flows and ascribed to the nearly bi-dimensional nature of these flows. 
Kraichnan showed that in 
two-dimensional (2D) turbulence, the energy is transferred from the forcing scale 
to larger scales due to the conservation of both energy and enstrophy by the 
inviscid dynamics \cite{kraichnan1967}. In a confined flow domain and without large scale friction, 
the energy accumulates at the largest possible scale, thus generating  coherent 
structures in the form of large scale vortices. 

It has been observed in laboratory experiments that the large scale circulation generated by forcing a nearly
$2$D flow at small scale can display random reversals \cite{sommeria}. The large scale velocity has a bimodal probability density function
(PDF) with two symmetric maxima related to the opposite signs of the large scale circulation. This regime bifurcates from another turbulent regime with a Gaussian velocity field with zero mean when the large scale friction is decreased. When the friction is decreased further, the reversals become less and less frequent and a condensed state with most of its kinetic energy in the large scale circulation is reached \cite{herault}. A similar sequence of transitions is observed in numerical simulations of the $2$D Navier-Stokes equation (NSE) with large scale friction and spatially periodic forcing \cite{mishra2015}. 

These transitions correspond to bifurcations of a mean flow on a strongly turbulent background for which no theoretical tool exists so far. 
We show in this Letter that the truncated Euler equation (TEE), {\emph i.e.} a finite set of ordinary differential equations (ODEs) for the amplitude 
of the large-scale modes without forcing and dissipation, can correctly describe these transitions. To wit, we compare the dynamical regimes observed in numerical simulations of the $2$D NSE with the ones obtained with the TEE when the characteristic scale $k_c^{-1}$ of the initial conditions is changed, with $k_c=\sqrt{\Omega/E}$ where $E$ is the kinetic energy of the flow and $\Omega$ is its enstrophy (integrated squared vorticity).

The dimensionless 2D NSE reads for an incompressible velocity field ${\bf u}=\nabla \times \psi$,
\begin{equation}
\frac{\partial \psi}{\partial t}-\frac{1}{\nabla^2}\{\psi,\nabla^2\psi\}= -\frac{1}{\rm Rh}\psi  + \frac{1}{\rm Re} \nabla^2 \psi 
+f_{\rm \psi} \, ,
\label{eq:NS_incom}
\end{equation}
where $\psi(x,y,t)$ is the stream function and $\{f,g\}=\partial_xf\partial_yg-\partial_xg\partial_yf$ is the usual Poisson bracket.
The first term on the right hand side represents the frictional force in the bottom boundary layer and 
$\nabla \times (f_{\rm \psi} \hat{\bf z})$ is the spatially periodic forcing, explicitly given by $f_{\rm \psi}=\frac{1}{144}\sin(6\,x)\sin(6\,y)$.

To model flow confinement we use free-slip boundary conditions; therefore, 
the stream-function can be Fourier expanded as
\begin{equation}\label{eq:sfbasis}
\psi(x,y)=\sum_{m,n}\hat{\psi}_{m,n}\sin(m x)\sin(n y).
\end{equation}
The non-dimensional parameters are the Reynolds number, 
${\rm Re}=UL/\nu$ and ${\rm Rh}=\tau U/L$, which represents the 
ratio of the inertial term to the friction on the bottom boundary. 
Here, $U$ is a characteristic large scale velocity, $L$ is the length 
of the square container, $\nu$ is the kinematic viscosity and $1/\tau$ is the damping rate related to the 
friction.  The above equation has been made dimensionless using the length scale $L$ and the velocity scale $U$.

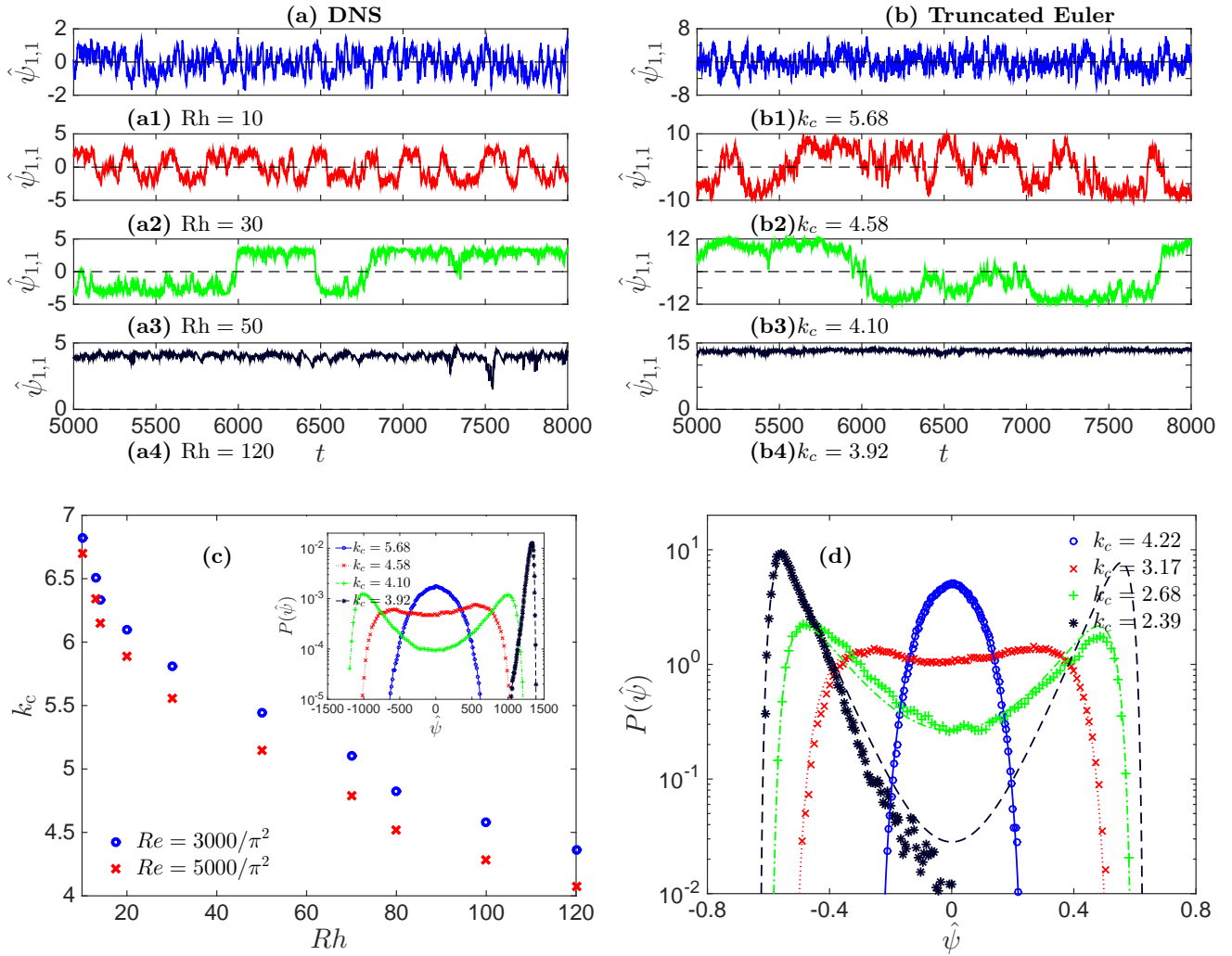
\begin{figure*}
\includegraphics[scale=0.45]{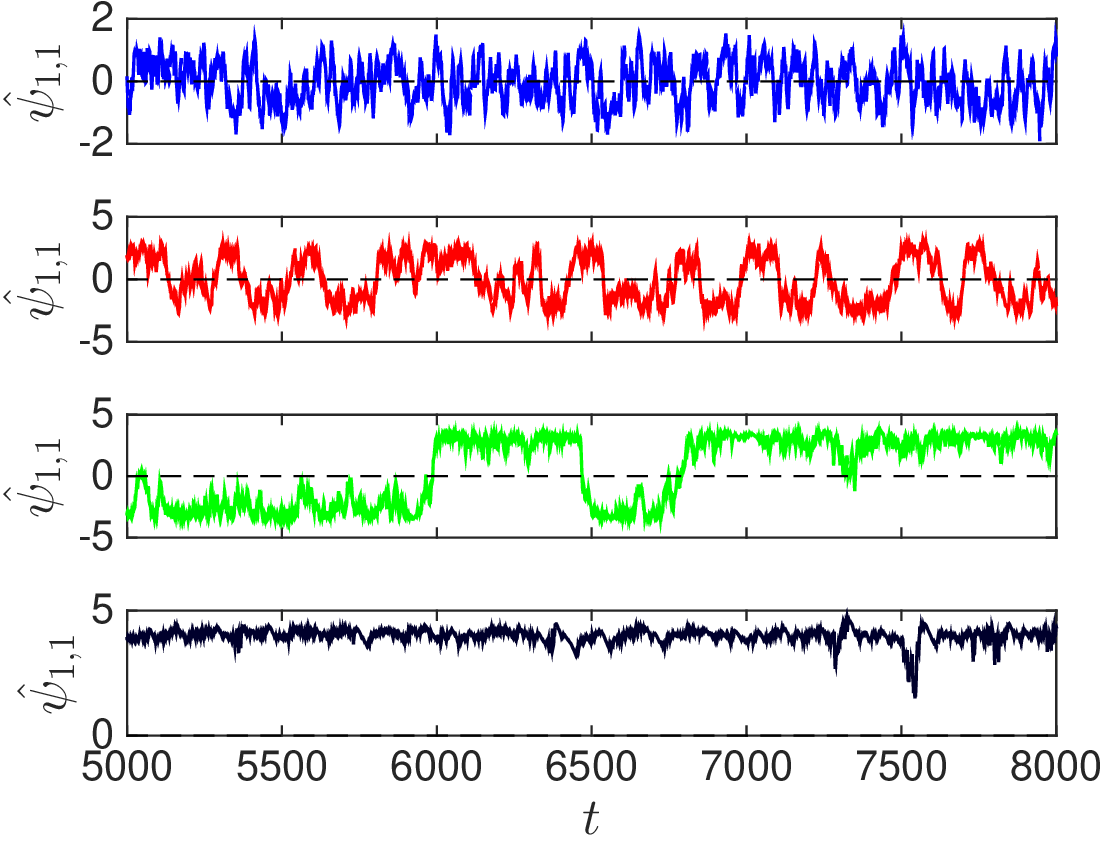}
\put(-125,182){\bf\small (a) DNS}
\put(-190,140){\bf\small(a1) ${\rm Rh}=10$}
\put(-190,97){\bf\small(a2) ${\rm Rh}=30$}
\put(-190,55){\bf\small(a3) ${\rm Rh}=50$}
\put(-190,4){\bf\small(a4) ${\rm Rh}=120$}
\hspace{0.25cm}
\includegraphics[scale=0.45]{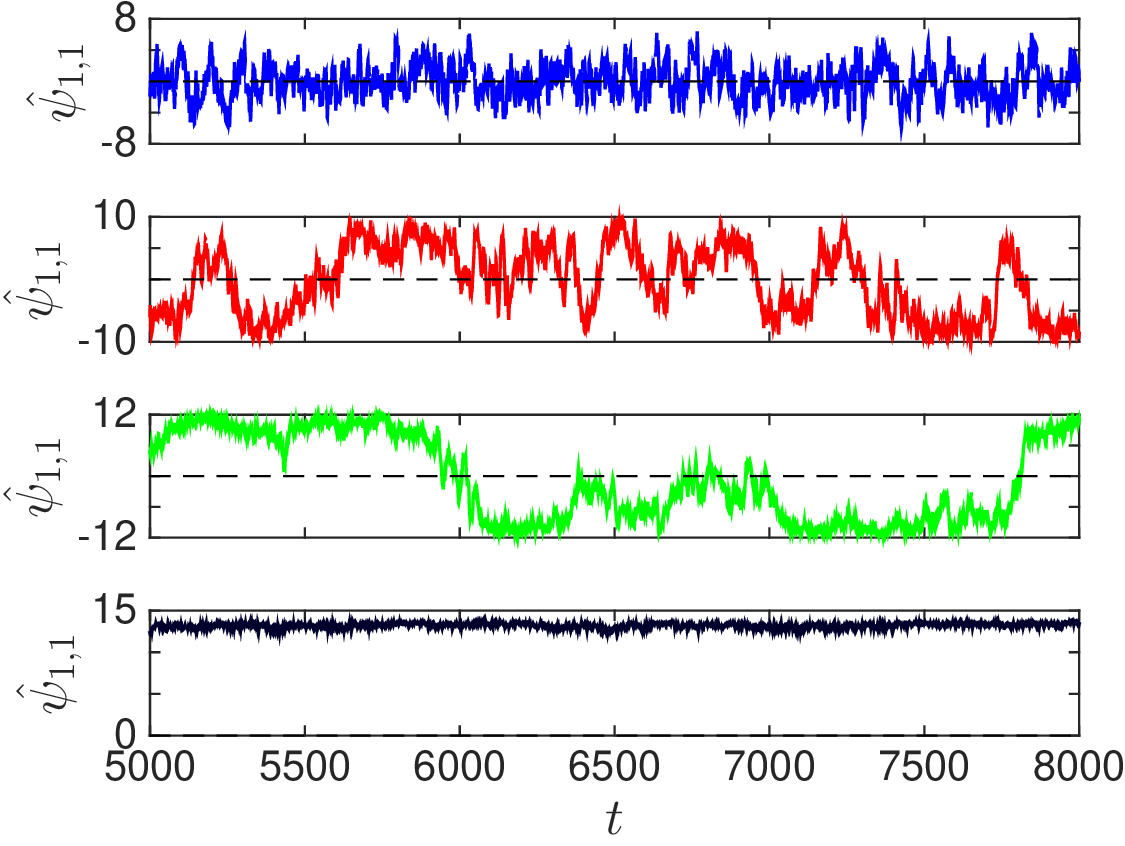}
\put(-135,182){\bf\small (b) Truncated Euler}
\put(-190,140){\bf\small(b1)$k_c=5.68$}
\put(-190,97){\bf\small(b2)$k_c=4.58$}
\put(-190,55){\bf\small(b3)$k_c=4.10$}
\put(-190,4){\bf\small(b4)$k_c=3.92$}
\\
\vspace{0.5cm}
\includegraphics[scale=0.45]{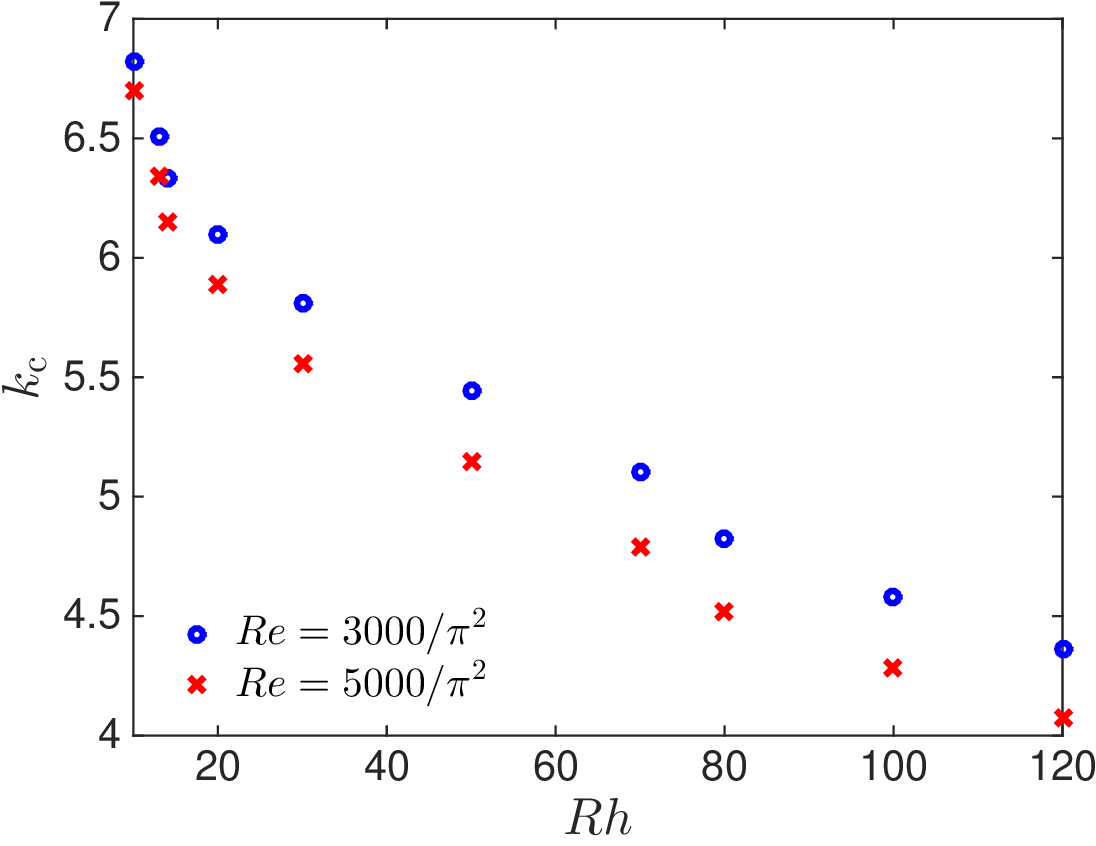}
\put(-160,160){\bf\small(c)}
\put(-130,89){\includegraphics[scale=0.21]{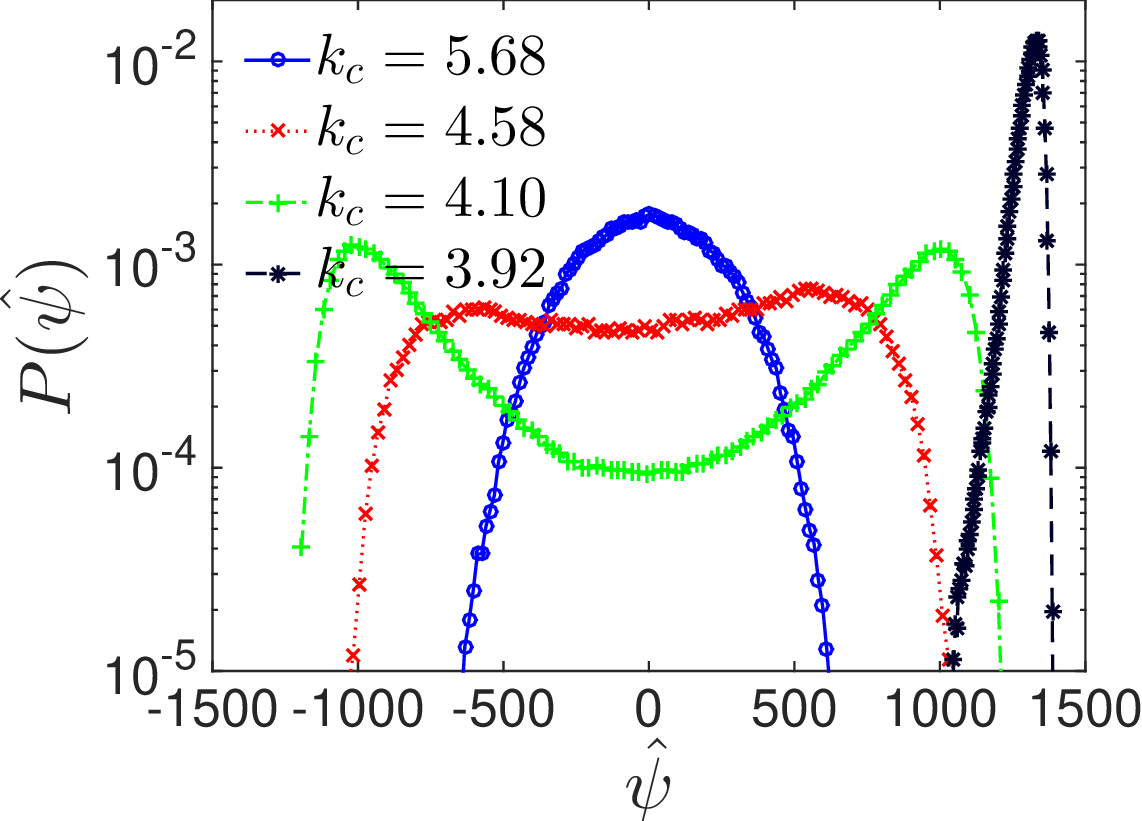}}
\hspace{0.25cm}
\includegraphics[scale=0.45]{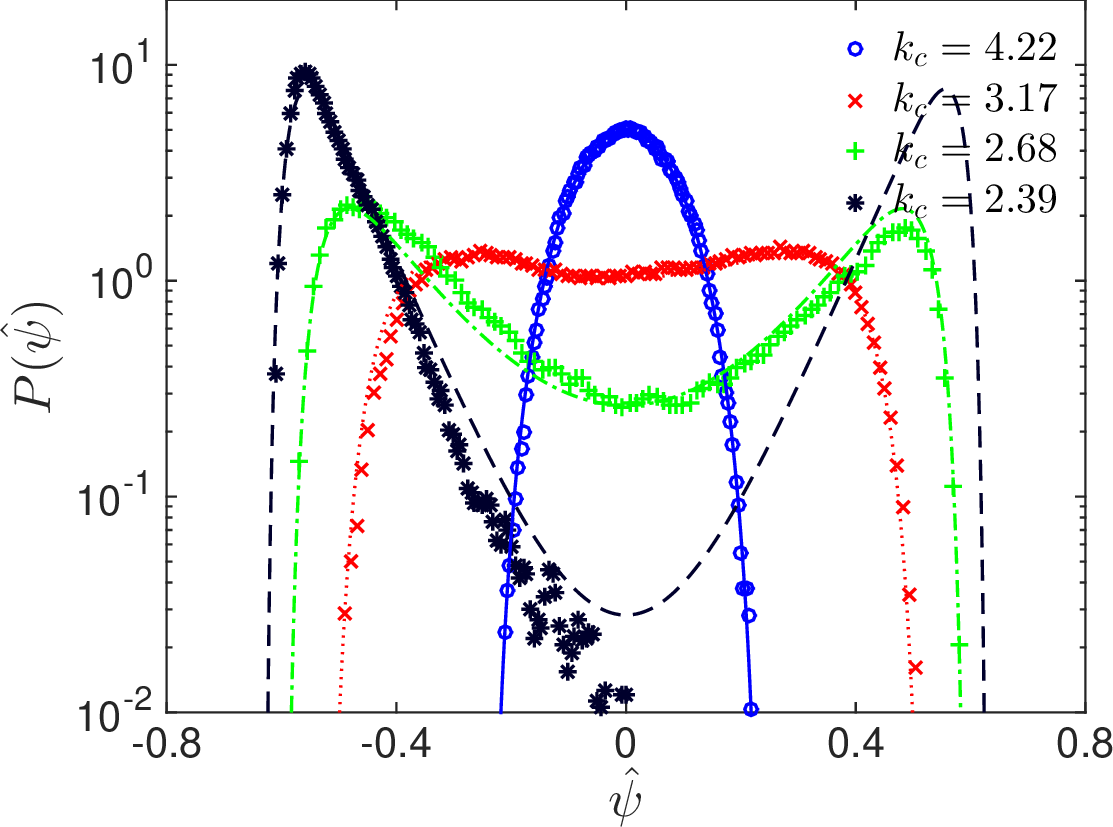}
\put(-160,160){\bf\small(d)}
\caption{(Color online) Flow transitions: times series of $\hat{\psi}_{1,1}$ obtained from the DNS of the
(a) Navier-Stokes equation (NSE) for different values of ${\rm Rh}$ and ${\rm Re}=5000/\pi^2$,
(b) truncated-Euler equation (TEE) for different values of $k_c$. 
$\hat{\psi}_{1,1}^{NSE}$ and $\hat{\psi}_{1,1}^{TEE}$ have been divided by
$10^4$ and $10^2$, respectively.
(c) Plot of $k_c=\sqrt{\Omega/E}$ versus ${\rm Rh}$ from the DNS of NSE for two different
Reynold's numbers ${\rm Re}=3000/\pi^2$ (blue circles) and ${\rm Re}=5000/\pi^2$ (red crosses).
Inset: Semilogy plots of the PDFs of $\hat{\psi}_{1,1}^{TEE}$ for different values of $k_c$. 
(d) Semilogy plots of the PDFs of $\hat{\psi}_{1,1}$ for different values of $k_c$ 
obtained from the finite-mode minimal model based on TEE; the lines on top
of these PDFs indicate the estimation from our analytical method.
}
\label{fig:1}
\end{figure*}

We perform direct numerical simulations (DNS) of Eq.~\eqref{eq:NS_incom},
using standard pseudospectral methods \cite{Got-Ors}
with $N_c^2$ collocation points and $2/3$ circular dealiasing:
$k_{\rm max}=N_c/3$.
Time stepping is performed with a second-order, exponential time differencing 
Runge-Kutta method~\cite{cox2002etd}. DNSs of the NSE~(\ref{eq:NS_incom}) are carried out 
for ${\rm Re}=3000/\pi^2$ and ${\rm Re}=5000/\pi^2$ with $N_c=256$.
Very long time integration is needed to accumulate 
reliable statistics for the reversals, which become rare with increase in ${\rm Rh}$ (see below).

Direct time recordings of the amplitude of the lowest wave number mode of the stream 
function, $\hat{\psi}_{1,1}$,
are displayed in Fig.~\ref{fig:1}(a) for ${\rm Re}=5000/\pi^2$ and different values of ${\rm Rh}$. For ${\rm Rh}=10$, the amplitude of the large scale circulation fluctuates around zero and its PDF is Gaussian (not shown). When ${\rm Rh}$ is increased, a first transition is observed within 
the turbulent regime and can be characterized by a change of the shape of the PDF that becomes bimodal. $\hat{\psi}_{1,1}$ fluctuates around two non zero most probable opposite values and displays random transitions between these two states (see Fig.~\ref{fig:1}(a2)). This corresponds to random reversals of the  large scale circulation on a turbulent background. The mean waiting time between successive reversals increases with ${\rm Rh}$ (see Fig.~\ref{fig:1}(a3)) and finally a large scale circulation with a given sign becomes the dominant flow component (see Fig.~\ref{fig:1}(a4)).  This is the condensed state described by Kraichnan \cite{kraichnan1967,kraichnan1975}. The regime with random reversals of the large scale circulation is therefore located in parameter space between the condensed states and the turbulent regime with Gaussian velocity PDFs as observed in experiments \cite{herault}.

We now consider the approach, introduced by Lee \cite{lee1952} 
and developed by Kraichnan \cite{kraichnan1967,kraichnan1975} that relies on 
the 2D TEE. They showed that the Euler equation, 
truncated between a minimum and a maximum wave number, gives a set of ODEs for the amplitudes of the modes that follow a 
Liouville theorem \cite{lee1952}. For 2D flows, the kinetic energy $E$ 
and the enstrophy $\Omega$ (integrated squared vorticity) are conserved; 
therefore, the Boltzmann-Gibbs canonical equilibrium distribution is of the form   
$\mathcal P = Z^{-1} \exp(-\alpha E -\beta \Omega)$, where $Z$ is the 
partition function and $\alpha$ and $\beta$ can be seen as inverse 
temperatures, determined by the total energy and enstrophy. 
Using this formalism, Kraichnan \cite{kraichnan1967,kraichnan1975} derived the absolute 
equilibria of the energy spectrum $E(k)$ 
and showed the existence of different regimes depending 
on the values of $\alpha$ and $\beta$. On the other hand, microcanonical distributions are defined by 
$\delta(E-E_0)\delta(\Omega-\Omega_0)$, where $E_0$ and $\Omega_0$ are respectively the energy and enstrophy
of the initial conditions, and should be used to compute the PDFs in the reversal and 
condensed state (see below).

The TEE is obtained by performing a circular Galerkin truncation at wave-number $k_{\rm max}$
of the incompressible, Euler equation 
$\frac{\partial \psi}{\partial t}-\frac{1}{\nabla^2}\{\psi,\nabla^2\psi\}=0$,
which is Eq.~\eqref{eq:NS_incom} without forcing or dissipation.
The TEE in spectral space reads
\begin{equation}
\frac{\partial \psi_{\bf k}}{\partial t}=\frac{1}{k^2}\sum_{\bf p,q} ( {\bf p}\times{\bf q})q^2
\psi_{\bf p} \psi_{\bf q} \delta_{{\bf k},{\bf p+q}} \label{eq:TEE}
\end{equation}
with $\delta_{{\bf k},{\bf r}}$ the Kronecker delta and with Fourier modes satisfying $\psi_{\bf k}=0$ if $|{\bf k}|\ge k_{\rm max}$.
Note that, because of the free-slip boundary conditions Eq.~\eqref{eq:sfbasis}, the Fourier modes $\psi_{\bf k}$ are real numbers.
This truncated system exactly conserves the quadratic invariants, energy and enstrophy, given in Fourier space by 
$E=\frac{1}{2}\sum_{\bf k}|{\bf u}_{\bf k}|^2$ and $\Omega=\frac{1}{2}\sum_{\bf k}k^2|{\bf u}_{\bf k}|^2$.

For TEE we take $k_{\rm max}$ as
a free parameter and the same stream function expansion as that used for the NS Eq.~\eqref{eq:sfbasis}, 
thus the numerical integration method is the same as the one described above for the NSE.
In both cases, the minimum wavenumber is $k_{\rm min}=\sqrt{2}$.
We use an initial velocity field with an energy spectra $E(k) = k/(\alpha+\beta k^2)$, where by varying $\alpha$ and $\beta$ we
can obtain different flow regimes in accordance with the Kraichnan's absolute equilibrium predictions. 
We introduce a wave-number $k_c$ given by
$k_c^2=\Omega/E$,
which acts as an important control parameter of the system.

We next consider the results obtained using the TEE~\eqref{eq:TEE} with $k_{\rm max}=k_f$ 
(the NSE forcing wavenumber)
and initial conditions with different values of $k_c$. Figure~\ref{fig:1}(b) shows the 
transitions between different turbulent regimes when $k_c$ is decreased. 
The corresponding PDFs of $\hat{\psi}_{1,1}$ obtained for different values of $k_c$ 
are displayed in the inset of Fig.~\ref{fig:1}(c). We observe the transition from 
Gaussian to bimodal PDF when $k_c$ is decreased and then the transition to 
the condensed regime with a given sign of the large scale circulation. 

For NSE at large ${\rm Re}$, the effect of the large scale friction is to stop the 
inverse cascade before reaching the scale of the flow domain. ${\rm Rh}$ 
thus determines the largest scale of the flow that we can define using the
wave number $k_c^2=\Omega/E$. $k_c$ is displayed in Fig.~\ref{fig:1}(c) for two values 
of ${\rm Re}$. It weakly depends on ${\rm Re}$ and monotonously decreases 
when ${\rm Rh}$ is increased. When ${\rm Rh}$ is
large (small friction), the kinetic energy accumulates at the scale of the 
flow domain and the condensed state is obtained. 

Although, we do not have a quantitative agreement between the transition 
values for $k_c$ for the TEE and ${\rm Rh}$ for the 
NSE when using the relation between $k_c$ and ${\rm Rh}$ displayed in
Fig.~\ref{fig:1}(c), the same sequence of transitions is observed in both cases. 
Figure~\ref{fig:1}(d) shows 
that we keep this qualitative agreement when the truncation is lower,  
$k_{\rm max}= 2\sqrt{5}$. This truncation leads to only $13$ ODEs
for the amplitudes of the large scale modes (see Supplemental Material~\cite{suppmat}). 
It is remarkable that this set of equations correctly describes the transitions
observed between the different turbulent regimes observed in direct numerical simulations 
and experiments.

The TEE model~\eqref{eq:TEE} is a finite number of quadratic nonlinear ODEs for real 
variables $y_i$ (see the remark following Eq.~\eqref{eq:TEE} about the amplitudes of the 
Fourier modes noted $y_i$ hereafter to simplify the notations)
that conserve both the energy $E(t)=\sum_{j=0}^n b_E(j)y_j^2$ and the enstrophy $\Omega(t)=\sum_{j=0}^n b_{\Omega}(j)y_j^2$ (see Supplemental Material~\cite{suppmat}).
By making use of the identities
\begin{equation}
\begin{split}
&(2 \pi)^2 \delta(b_Ey^2-E)\delta(b_{\Omega} y^2-\Omega)\\
&=\int_{-\infty}^\infty dp_E dp_{\Omega} e^{ip_E (b_E y^2-E)+ i p_{\Omega} (b_{\Omega} y^2-\Omega)}
\end{split}
\end{equation}
and
\begin{equation}
\begin{split}
\int_{-\infty}^\infty  e^{i (p_E b_E + p_{\Omega} b_{\Omega}) y^2}dy=\frac{\sqrt{\pi }}{\sqrt{-i (b_E p_E +b_{\Omega} p_{\Omega})}}
\end{split}
\end{equation}
we can write the total microcanonical phase space volume 
\begin{equation}
{\mathcal V}=\int\prod_{j=0}^n d y_j \delta\left(\sum_{j=0}^n b_E(j)y_j^2-E \right)\delta\left(\sum_{j=0}^n b_{\Omega}(j)y_j^2-\Omega\right)\label{eq:vol}
\end{equation} 
as
\begin{equation}
{\mathcal V}=\int_{-\infty}^\infty dp_E dp_{\Omega} e^{L_{\mathcal V}(p_E, p_{\Omega})}\label{eq:int}
\end{equation}
with
\begin{equation}
\begin{split}
L_{\mathcal V} &= \sum _{j=0}^n \log \left(\frac{\sqrt{\pi }}{\sqrt{-i (p_E b_E(j)+p_{\Omega}
   b_{\Omega}(j))}}\right)\\
 &+  i p_E (-E) + i p_{\Omega} (-\Omega)- 2 \log(2 \pi)
 \end{split}
 \end{equation}
 Using the steepest descent method \cite{Zinn1995,benderorszag} on the integral Eq.~\eqref{eq:int}, 
 the expression
\begin{equation}
\begin{split}
{\mathcal V}(p_E, p_{\Omega}) = 2\,\pi\,e^{L_{\mathcal V}}
 \Bigl(\det(\partial^2L_{\mathcal V}/\partial_{p_E}\partial_{p_{\Omega}})\Bigr)^{-1/2}\label{eq:res}
\end{split}
\end{equation}
furnishes an explicit parametric expression for ${\mathcal V}$ at the saddle-point $(p_E,p_{\Omega})$  \footnote{Purely imaginary values for $(p_E,p_{\Omega})$ parametrize real values of $(E,\Omega)$)}
that corresponds to values of energy and enstrophy given by the saddle conditions
\begin{eqnarray}
E(p_E, p_{\Omega})&=&i \sum _{j=0}^n \frac{b_E(j)}{2 (p_E b_E(j)+p_{\Omega} b_{\Omega}(j))} \nonumber \\
\Omega(p_E, p_{\Omega})&=&i \sum _{j=0}^n \frac{b_{\Omega}(j)}{2 (p_E b_E(j)+p_{\Omega} b_{\Omega}(j))}\label{eq:saddle}
\end{eqnarray}
and thus to $k_c^2=\Omega/E$.

We can estimate the same way the phase space volume for a fixed value of $y_0$ by 
retracing the steps from Eq.~\eqref{eq:vol} to \eqref{eq:res}, but with 
the product and sums going from $1$ to $n$ instead of $0$ to $n$.  
By combining these parametric representations, we obtain an explicit expression for the 
normalized PDF of $y_0$ that is shown in Fig.\ref{fig:1}(d) and displays a good agreement 
with the numerical results \footnote{We have checked (data not shown) that the 
steepest-descent estimates correctly represent microcanonical Monte-Carlo results, 
even in the highly bimodal regime}.

Note that canonical distributions with quadratic invariants are gaussian. When there is 
condensation of energy at large scale, only a few modes are present and then the canonical 
distribution has no reason to reproduce the microcanonical distribution results
~\cite{KellsOrszag1978}. Indeed, $k_c^2=\Omega/E<5$ with 
$\Omega=(1^2+1^2)^2 y_0^2+(1^2+2^2)^2 y_1^2+\dots$ and 
$E=(1^2+1^2) y_0^2+(1^2+2^2) y_1^2+\dots$ implies that $y_0\neq0$. 
Thus the microcanonical PDF of $y_0$  has to obey $p(0)=0$ for $k_c^2<5$ 
which forbids reversals of the large scale circulation.
This represents ergodicity breakdown. Our above results show that this breakdown 
is preceded by an ergodicity delay, in the sense that 
$p(0)$ becomes very small \cite{Mininni2014,Matthaeus2011,Shebalin1998}.\\

\begin{figure}
\includegraphics[scale=0.45]{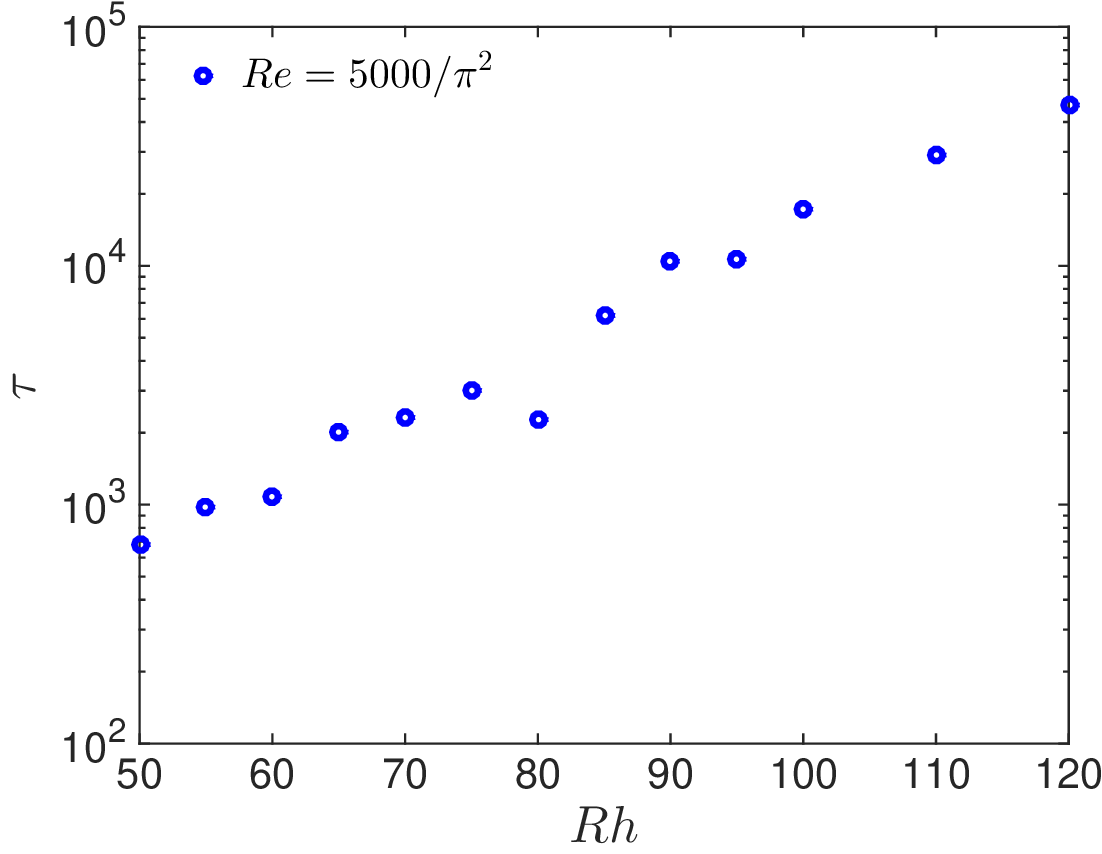}
\put(-95,30){\includegraphics[scale=0.15]{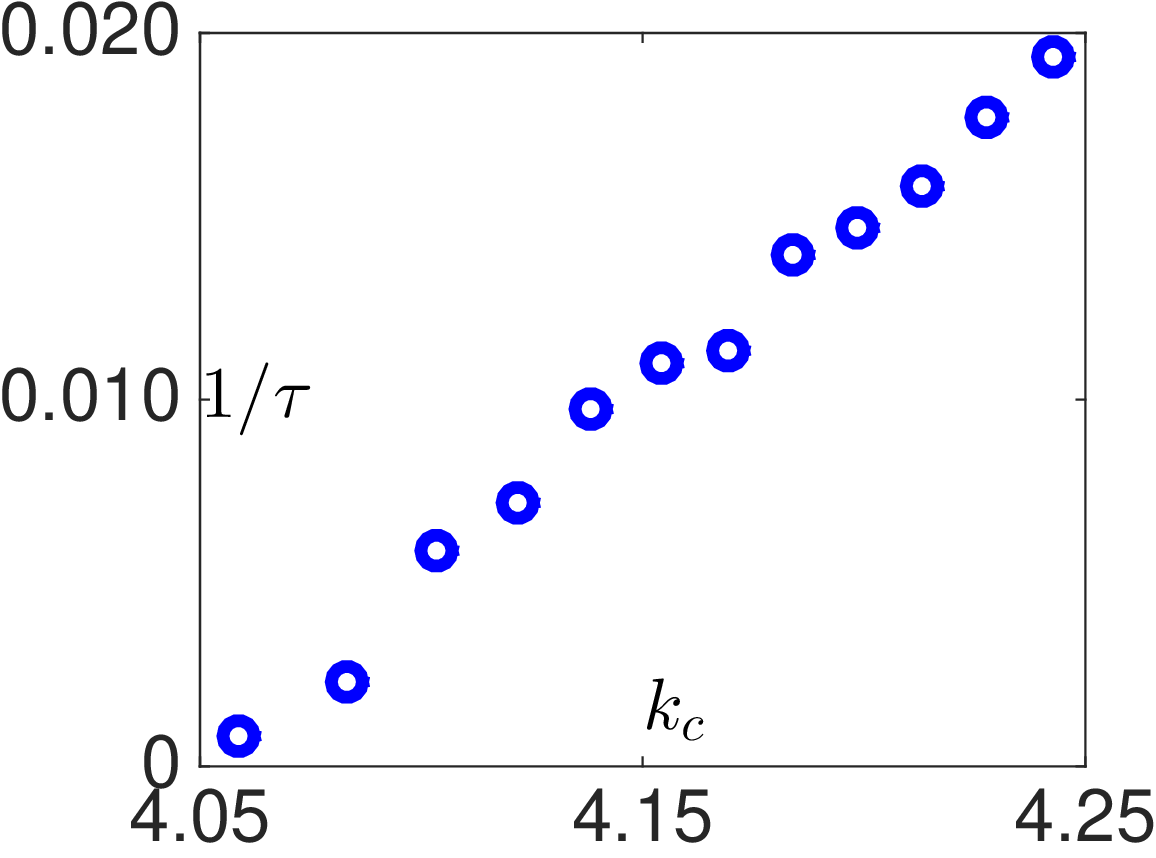}}
\caption{ (Color online) Reversals: 
Semilogy plot of the mean waiting time $\tau$ between successive reversals versus 
${\rm Rh}$, obtained from our DNSs of NSE for ${\rm Re}=5000/\pi^2$ 
(blue circles).
Inset: Plot of the reversal frequency $1/\tau$ versus $k_c$ from the DNSs of TEE;
it shows that the reversal frequency decreases linearly with $k_c$ with a critical
$k_c^*\simeq 4.06$, below which the reversals are not observed for the integration time.
}
\label{fig:2}
\end{figure}

We now consider in more detail the regime with random reversals of the 
large scale circulation and its transition
to the condensed regime for which the flow 
no longer explores the whole phase space, keeping a given sign of 
the large scale circulation. As shown above (compare Fig.~\ref{fig:1}(a) and (b)), 
the mean waiting time $\tau$ between successive reversals increases when 
${\rm Rh}$ is increased in the NSE, respectively $k_c$ is decreased in the TEE. 
However, the divergence of $\tau$ does not follow the same law for the 
NSE and the TEE. Figure~\ref{fig:2}(a) shows an exponential increase
of $\tau$ with ${\rm Rh}$ in the NSE, whereas a fit of the form 
$\tau \propto (k_c - k_c^*)^{-1}$ with $k_c^*\simeq 4.06$
is observed in the TEE (see Fig.~\ref{fig:2}(b)). The later result is expected since 
there exists a critical value of $k_c$ below which reversals are not possible 
in order to fulfill the conservation of both $E$ and $\Omega$. We thus expect 
that $\tau$ becomes infinite for a finite value of $k_c$. A similar trend is 
not observed in the NSE for $\tau$ versus ${\rm Rh}$. This cannot be 
explained using the relation between $k_c$ and ${\rm Rh}$ displayed in 
Fig.~\ref{fig:1}(c) that is roughly linear close to the transition
to the condensed regime. In contrast to the TEE, the NSE does not 
involve conserved quantities that prevent reversals,
even when ${\rm Rh}$ is large. In addition, all the modes above $k_f$ that 
are suppressed in the TEE can act as an additional source of noise in the 
NSE and trigger reversals.\\

Although it can be expected that viscous dissipation is negligible for the dynamics of large scales,
it is remarkable that taking into account the effect of large scale friction by selecting the value of $\Omega/E$ in the
initial conditions of the TEE is enough to describe the bifurcations of the large scale flow using a small number of 
modes governed by the Euler equation. Thus, one discards the huge number of degrees of freedom related to
small scale turbulent fluctuations. In addition, equilibrium statistical mechanics, using the microcanonical distribution related to the TEE, 
correctly describes the PDF of the large scale velocity in the different turbulent regimes.  Transitions between different
mean flows are widely observed in turbulent regimes, the most famous example being the drag crisis for which the wake of a sphere
becomes narrower. Using the Navier-Stokes equation with noisy forcing \cite{bouchet2009} is a way to describe this type of transitions.
The TEE as presented here, can provide a new method to describe the dynamics of large scales in turbulence
and to model a bifurcation of the mean flow on a strongly turbulent background.\\

\begin{acknowledgments}
We thank Francois P\'etr\'elis for useful discussions.
Support of the Indo-French Centre for the Promotion of Advanced Research (IFCPAR/CEFIPRA) contract 4904-A is acknowledged. 
This work was granted access to the HPC ressources of MesoPSL financed by the Region Ile de France and the project Equip@Meso 
(Reference No. ANR-10-EQPX-29-01) of the programme Investissements d'Avenir supervised by the Agence Nationale pour la Recherche.
VS acknowledges supported from EuHIT - European High- performance Infrastructure in Turbulence, which is funded by the European Commission Framework Program 7 (Grant No. 312778). 
\end{acknowledgments}


%

\section{Supplemental Material}

We give here explicitly the set of thirteen ordinary differential 
equations (ODEs) for the amplitudes of the Fourier modes $\hat{\psi}_{m,n}$
that define the Truncated Euler Equation (TEE) in the case $k_{\rm max}=2\sqrt{5}$. 
Note that $\hat{\psi}_{m,n}$ are real numbers because of the free-slip boundary conditions (see text).
\begin{equation}
\begin{split}
\frac{d \hat{\psi}_{11}}{d t}&=\frac{1}{2}
\biggl[
\bigl(2 \hat{\psi}_{12}+5\hat{\psi}_{14}\bigr)\hat{\psi}_{23}
+\hat{\psi}_{13} \bigl(2 \hat{\psi}_{22}+5 \hat{\psi}_{24}\bigr)\\
&-2 \hat{\psi}_{22}\hat{\psi}_{31} -\hat{\psi}_{32}\bigl(2\hat{\psi}_{21}+5\hat{\psi}_{41}\bigr)\\
&+3\hat{\psi}_{33}\bigl(\hat{\psi}_{24}-\hat{\psi}_{42}\bigr)
-5\hat{\psi}_{31}\hat{\psi}_{42}
\biggr]
\end{split}
\end{equation}

\begin{equation}
\begin{split}
\frac{d \hat{\psi}_{12}}{d t} &=
 \frac{1}{20} \biggl[
25 \hat{\psi}_{13}\hat{\psi}_{21} + 54 \hat{\psi}_{14}\hat{\psi}_{22} \\
&+\hat{\psi}_{11} \bigl(9 \hat{\psi}_{21} - 11 \hat{\psi}_{23}\bigr) + 25 \hat{\psi}_{21} \hat{\psi}_{31} \\
&+ 21 \hat{\psi}_{23}\hat{\psi}_{31} + 56 \hat{\psi}_{24}\hat{\psi}_{32} - 39 \hat{\psi}_{21}\hat{\psi}_{33} \\
&+   49 \hat{\psi}_{31} \hat{\psi}_{41} + 9 \hat{\psi}_{33}\hat{\psi}_{41}
\biggr]
\end{split}
\end{equation}

\begin{equation}
\begin{split}
\frac{d \hat{\psi}_{21}}{d t} &=
\frac{1}{20}\biggl[
-49\hat{\psi}_{13}\hat{\psi}_{14} - \hat{\psi}_{11} \bigl(9\hat{\psi}_{12} - 11\hat{\psi}_{32}\bigr)\\
& -    21\hat{\psi}_{13}\hat{\psi}_{32} - \hat{\psi}_{12}\bigl(25\hat{\psi}_{13} + 25\hat{\psi}_{31} - 39\hat{\psi}_{33}\bigr)\\
& -9\hat{\psi}_{14}\hat{\psi}_{33} - 54\hat{\psi}_{22}\hat{\psi}_{41} - 56\hat{\psi}_{23}\hat{\psi}_{42}
\biggr]
\end{split}
\end{equation}

\begin{equation}
\begin{split}
\frac{d \hat{\psi}_{22}}{d t} &=
\frac{1}{4}\biggl[-9 \hat{\psi}_{12}\hat{\psi}_{14} - 4\hat{\psi}_{11}\bigl(\hat{\psi}_{13} - \hat{\psi}_{31}\bigr) \\
&+ 5\hat{\psi}_{14}\hat{\psi}_{32} + 9\hat{\psi}_{21}\hat{\psi}_{41} - 5\hat{\psi}_{23}\hat{\psi}_{41}
\biggr]
\end{split}
\end{equation}

\begin{equation}
\begin{split}
\frac{d \hat{\psi}_{13}}{d t} &=
\frac{1}{10} \biggl[\hat{\psi}_{11} (6\hat{\psi}_{22} - 9\hat{\psi}_{24}) + 7\hat{\psi}_{21}\bigl(3\hat{\psi}_{14} + 2\hat{\psi}_{32}\bigr) \\
&+ 11\hat{\psi}_{32}\hat{\psi}_{41} + \hat{\psi}_{31}\bigl(4\hat{\psi}_{22} + 25(\hat{\psi}_{24} + \hat{\psi}_{42})\bigr)
\biggr]
\end{split}
\end{equation}

\begin{equation}
\begin{split}
\frac{d \hat{\psi}_{31}}{d t} &=
\frac{1}{10} \biggl[-14\hat{\psi}_{12}\hat{\psi}_{23} - 11\hat{\psi}_{14}\hat{\psi}_{23} - 21\hat{\psi}_{12}\hat{\psi}_{41} \\
&-\hat{\psi}_{11}\bigl(6\hat{\psi}_{22} - 9\hat{\psi}_{42}\bigr) \\
&-\hat{\psi}_{13}\bigl(4\hat{\psi}_{22} + 25(\hat{\psi}_{24} + \hat{\psi}_{42})\bigr)
\biggr]
\end{split}
\end{equation}

\begin{equation}
\begin{split}
\frac{d \hat{\psi}_{23}}{d t} &=
\frac{1}{52}\biggl[35\hat{\psi}_{12}\hat{\psi}_{31} + 77\hat{\psi}_{14}\hat{\psi}_{31} \\
&+   \hat{\psi}_{11}\bigl(3\hat{\psi}_{12} - 75\hat{\psi}_{14} + 55\hat{\psi}_{32}\bigr) \\
&+ 90\hat{\psi}_{22}\hat{\psi}_{41} + 42\hat{\psi}_{24}\hat{\psi}_{41} + 120\hat{\psi}_{21}\hat{\psi}_{42}
\biggr]
\end{split}
\end{equation}

\begin{equation}
\begin{split}
\frac{d \hat{\psi}_{32}}{d t} &=
\frac{1}{52}\biggl[-\hat{\psi}_{11}\bigl(3\hat{\psi}_{21} + 55\hat{\psi}_{23} - 75\hat{\psi}_{41}\bigr)\\
& -  7 \hat{\psi}_{13}\bigl(5\hat{\psi}_{21} + 11\hat{\psi}_{41}\bigr) \\
&-6 \bigl(20\hat{\psi}_{12}\hat{\psi}_{24} + \hat{\psi}_{14}(15\hat{\psi}_{22} + 7\hat{\psi}_{42})\bigr)
\biggr]
\end{split}
\end{equation}

\begin{equation}
\begin{split}
\frac{d \hat{\psi}_{14}}{d t} &=
\frac{1}{68}\biggl[-35\hat{\psi}_{13}\hat{\psi}_{21} + 18\hat{\psi}_{12}\hat{\psi}_{22} + 55\hat{\psi}_{11}\hat{\psi}_{23}\\
& -33\hat{\psi}_{23}\hat{\psi}_{31} + 50\hat{\psi}_{22}\hat{\psi}_{32} \\
&+ 117\hat{\psi}_{21}\hat{\psi}_{33} 
- 15\hat{\psi}_{33}\hat{\psi}_{41} + 98\hat{\psi}_{32}\hat{\psi}_{42}
\biggr]
\end{split}
\end{equation}

\begin{equation}
\begin{split}
\frac{d \hat{\psi}_{41}}{d t} &=
\frac{1}{68}\biggl[-18\hat{\psi}_{21}\hat{\psi}_{22} - 50\hat{\psi}_{22}\hat{\psi}_{23} - 98\hat{\psi}_{23}\hat{\psi}_{24}\\
& + 35\hat{\psi}_{12}\hat{\psi}_{31} - 55\hat{\psi}_{11}\hat{\psi}_{32} \\
&+ 33\hat{\psi}_{13}\hat{\psi}_{32} - 117\hat{\psi}_{12}\hat{\psi}_{33} + 15\hat{\psi}_{14}\hat{\psi}_{33}
\biggr]
\end{split}
\end{equation}

\begin{equation}
\begin{split}
\frac{d \hat{\psi}_{24}}{d t} & =
\frac{1}{10}\biggl[8 \hat{\psi}_{12}\hat{\psi}_{32} + 2\hat{\psi}_{11}\bigl(\hat{\psi}_{13} + 6\hat{\psi}_{33}\bigr) \\
& +7\hat{\psi}_{23}\hat{\psi}_{41} + 18\hat{\psi}_{22}\hat{\psi}_{42}
\biggr]
\end{split}
\end{equation}

\begin{equation}
\begin{split}
\frac{d \hat{\psi}_{42}}{d t} &=
\frac{1}{10} \biggl[-8 \hat{\psi}_{21}\hat{\psi}_{23} - 18\hat{\psi}_{22}\hat{\psi}_{24} - 2\hat{\psi}_{11}\hat{\psi}_{31} \\
&-7\hat{\psi}_{14}\hat{\psi}_{32} - 12\hat{\psi}_{11}\hat{\psi}_{33}
\biggr]
\end{split}
\end{equation}

\begin{equation}
\begin{split}
\frac{d \hat{\psi}_{33}}{d t} &=
-\frac{3}{2}\biggl[\hat{\psi}_{14}\hat{\psi}_{21} - \hat{\psi}_{12}\hat{\psi}_{41} \\
&+ \hat{\psi}_{11}\bigl(\hat{\psi}_{24} - \hat{\psi}_{42}\bigr)
\biggr]
\end{split}
\end{equation}

The dynamical evolution of the above set of ODEs conserves the total energy
and enstrophy, which are given by
\begin{equation}
\begin{split}
E &= (1^2 + 1^2)\hat{\psi}_{11}^2 + (1^2 + 2^2)\hat{\psi}_{12}^2 + (2^2 + 1^2)\hat{\psi}_{21}^2 \\
&+ (2^2 + 2^2)\hat{\psi}_{22}^2 + (3^2 + 1^2)\hat{\psi}_{31}^2 + (1^2 + 3^2)\hat{\psi}_{13}^2 \\
&+ (2^2 + 3^2)\hat{\psi}_{23}^2 + (3^2 + 2^2)\hat{\psi}_{32}^2 + (1^2 + 4^2)\hat{\psi}_{14}^2 \\
&+ (4^2 + 1^2)\hat{\psi}_{41}^2 + (2^2 + 4^2)\hat{\psi}_{24}^2 + (4^2 + 2^2)\hat{\psi}_{42}^2 \\
&+ (3^2 + 3^2)\hat{\psi}_{33}^2,
\end{split}
\end{equation}
and
\begin{equation}
\begin{split}
\Omega 
&= (1^2 + 1^2)^2\hat{\psi}_{11}^2 + (1^2 + 2^2)^2\hat{\psi}_{12}^2 + (2^2 + 1^2)^2\hat{\psi}_{21}^2 \\
&+ (2^2 + 2^2)^2\hat{\psi}_{22}^2 + (3^2 + 1^2)^2\hat{\psi}_{31}^2 + (1^2 + 3^2)^2\hat{\psi}_{13}^2 \\
&+ (2^2 + 3^2)^2\hat{\psi}_{23}^2 + (3^2 + 2^2)^2\hat{\psi}_{32}^2 + (1^2 + 4^2)^2\hat{\psi}_{14}^2 \\
&+ (4^2 + 1^2)^2\hat{\psi}_{41}^2 + (2^2 + 4^2)^2\hat{\psi}_{24}^2 + (4^2 + 2^2)^2\hat{\psi}_{42}^2 \\
&+ (3^2 + 3^2)^2\hat{\psi}_{33}^2.
\end{split}
\end{equation}

\end{document}